\documentclass[
	journal
]{IEEEtran}

\IEEEoverridecommandlockouts % This is to make the \thanks{} command visible in conference mode
\makeatletter
\def\markboth#1#2{\def\leftmark{\@IEEEcompsoconly{\sffamily}\MakeUppercase{\protect#1}}%
\def\rightmark{\@IEEEcompsoconly{\sffamily}\MakeUppercase{\protect#2}}}
\makeatother

\renewcommand{\markboth}[1]{\renewcommand{\leftmark}{#1}\renewcommand{\rightmark}{#1}}
\usepackage[amssymb]{SIunits}
\usepackage[cmex10]{amsmath}
\usepackage{amssymb}
\usepackage{amsthm}
\usepackage[latin1]{inputenc}
\usepackage{cite}
\usepackage{url}
\usepackage{glossaries}
\usepackage{graphicx}
\usepackage[hang,small,bf]{caption}
\usepackage[subrefformat=parens,labelformat=parens]{subfig}
\usepackage{bm}
\usepackage[usenames]{color}
\usepackage[lined,ruled,linesnumbered]{algorithm2e}
\usepackage{balance}

\newacronym{AWGN}{AWGN}{additive white Gaussian noise} 
\newacronym{CSA}{CSA}{coded slotted ALOHA} 
\newacronym{ABCSA}{B-CSA}{all-to-all broadcast CSA} 
\newacronym{SNR}{SNR}{signal-to-noise ratio} 
\newacronym{SINR}{SINR}{signal-to-interference-plus-noise ratio}
\newacronym{PLR}{PLR}{packet loss rate} 
\newacronym{UEP}{UEP}{unequal error protection} 
\newacronym{BS}{BS}{base station} 
\newacronym{LDPC}{LDPC}{low-density parity-check} 
\newacronym{VN}{VN}{variable node} 
\newacronym{CN}{CN}{check node} 
\newacronym{DE}{DE}{density evolution} 
\newacronym{MDS}{MDS}{maximum distance separable} 
\newacronym{BEC}{BEC}{binary erasure channel} 
\newacronym{PEC}{PEC}{packet erasure channel} 
\newacronym{DAMA}{DAMA}{demand assignment multiple access} 
\newacronym{CSMA}{CSMA-CA}{carrier sense multiple access with collision avoidance} 
\newacronym{VANET}{VANET}{vehicular ad hoc network} 
\newacronym{V2V}{V2V}{vehicle to vehicle} 
\newacronym{PHY}{PHY}{physical layer} 
\newacronym{MAC}{MAC}{medium access control} 
\newacronym{ARQ}{ARQ}{automatic repeat request} 
\newacronym{CDMA}{CDMA}{code division multiple access} 
\newacronym{TDMA}{TDMA}{time division multiple access}
\newacronym{CAM}{CAM}{cooperative awareness message}
\newacronym{DENM}{DENM}{decentralized environmental notification message}
\newacronym{ETSI}{ETSI}{European Telecommunications Standards Institute}
\newacronym{GPS}{GPS}{Global Positioning System}
\newacronym{DUEP}{DUEP}{double unequal error protection}
\newacronym{VC}{VC}{vehicular communication}
\newacronym{RU}{RU}{receiving user}
\newacronym{SIC}{SIC}{successive interference cancellation}

\newcommand{\figref}[1]{Fig.~\ref{#1}}

\newcommand{\tabref}[1]{Table~\ref{#1}}

\newcommand{\secref}[1]{Section~\ref{#1}}

\newcommand{\expect}[2]{\mathsf{E}_{#1}\left\{#2\right\}}

\newcommand{\setU}{\mathcal{U}}

\newcommand{\PEPt}{\bar{p}}

\newcommand{\Prss}{\rho}

\newcommand{\setS}{\mathcal{S}}

\newcommand{\maxd}{q}

\newcommand{\tframe}{t_{\text{frame}}}
\newcommand{\tslot}{t_{\text{slot}}}
\newcommand{\rdata}{r_{\text{data}}}
\newcommand{\packsize}{d_{\text{pack}}}
\newcommand{\tpream}{t_{\text{pream}}}
\newcommand{\tguard}{t_{\text{guard}}}
\newcommand{\tcsma}{t_{\text{csma}}}
\newcommand{\taifs}{t_{\text{aifs}}}
\newcommand{\tpack}{t_{\text{pack}}}

\newenvironment{DIFnomarkup}{}{}

\begin{document}

\begin{DIFnomarkup}

\title{All-to-all Broadcast for Vehicular Networks Based on Coded Slotted ALOHA}

\author{%
\IEEEauthorblockN{Mikhail~Ivanov, Fredrik Br\"{a}nnstr\"{o}m, Alexandre Graell i Amat, and Petar Popovski{\IEEEauthorrefmark{4}}\\}
\IEEEauthorblockA{Department of Signals and  Systems, Chalmers  University of Technology, Gothenburg, Sweden\\} \IEEEauthorblockA{ \IEEEauthorrefmark{4}Department of Electronic Systems, Aalborg University, Aalborg, Denmark\\ \emph{\{mikhail.ivanov, fredrik.brannstrom, alexandre.graell\}@chalmers.se, petarp@es.aau.dk}
}
\thanks{This research was supported by the Swedish Research Council, Sweden, under Grant No. 2011-5950, in part by the Ericsson's Research Foundation, Sweden, under Grant No. 556016-0680, in part by Chalmers Antenna Systems Excellence Center in the project `Antenna Systems for V2X Communication', and in part by the Danish Council for Independent Research within the Sapere Aude Research Leader program, Grant No. 11-105159.%The calculations were performed on resources provided by the Swedish National Infrastructure for Computing (SNIC) at C3SE.
}

}%

%\author{Mikhail~Ivanov, Fredrik Br\"{a}nnstr\"{o}m,~\IEEEmembership{Member,~IEEE}, Alexandre Graell i Amat,~\IEEEmembership{Senior Member,~IEEE},\\Petar Popovski,~\IEEEmembership{Senior Member,~IEEE}
%\thanks{This research was supported by the Swedish Research Council, Sweden, under Grant No. 2011-5950, and in part by the Ericsson's Research Foundation, Sweden, under Grant No. 556016-0680. The calculations were performed on resources provided by the Swedish National Infrastructure for Computing (SNIC) at C3SE.}
%\thanks{M. Ivanov, F. Br\"{a}nnstr\"{o}m, and A. Graell i Amat are with the Dept.~of Signals and Systems, Chalmers Univ.~of Technology, SE-41296 Gothenburg, Sweden (e-mail: \{mikhail.ivanov, fredrik.brannstrom, alexandre.graell\}@chalmers.se).}
%\thanks{Petar Popovski is with the Dept. of Electronic Systems, Aalborg University, 9220 Aalborg, Denmark (e-mail: petarp@es.aau.dk).}
%}

\maketitle

\end{DIFnomarkup}

\begin{abstract}
We propose an uncoordinated all-to-all broadcast protocol for periodic messages in vehicular networks based on coded slotted ALOHA (CSA). Unlike classical CSA, each user acts as both transmitter and receiver in a half-duplex mode. As in CSA, each user transmits its packet several times. The half-duplex mode gives rise to an interesting design trade-off: the more the user repeats its packet, the higher the probability that this packet is decoded by other users, but the lower the probability for this user to decode packets from others. We compare the proposed protocol with carrier sense multiple access with collision avoidance, currently adopted as a multiple access protocol for vehicular networks. The results show that the proposed protocol greatly increases the number of users in the network that reliably communicate with each other. We also provide analytical tools to predict the performance of the proposed protocol.
\end{abstract}

% Not needed for conference
%\begin{keywords} 
%	APSK constellation, nonlinear phase
%	noise, optical Kerr-effect, self-phase modulation. 
%\end{keywords}
% Redefine all acronyms that have been defined in the introduction
\glsresetall

\section{Introduction}\label{sec:intro}

Reliable \glspl{VC} is presently one of the most challenging problems of communication engineering. Its deployment will enable numerous applications, such as intelligent transportation systems and autonomous driving, as well as improve traffic safety. \glspl{VC} entails a number of challenges, such as high mobility networks with rapidly changing topologies and a large number of users, poor channel quality, and strict reliability and delay requirements. These challenges require new ideas and design at the \gls{PHY} and the \gls{MAC} layer.

The current status of \glspl{VC} is summarized in the standard~\cite{IEEE80211} and is usually referred to as 802.11p. The \gls{PHY} and the \gls{MAC} layer in~802.11p are based on the Wi-Fi protocol that works well for low mobility networks. In the context of \glspl{VC}, the \gls{PHY} is mainly criticized for not being able to cope with time-varying channels~\cite{Zemen12}. 
802.11p uses \gls{CSMA} as the \gls{MAC} protocol. It does not require any coordination and is shown to work well for networks with a small number of users, large amounts of information to be transmitted at each user, and no delay constraints. Under these conditions, \gls{CSMA} can provide large throughputs~\cite{Bianchi00}. In \glspl{VANET}, however, the number of users is large, the amount of information to be transmitted is rather small, and there are hard deadlines for accessing the channel. In such scenarios, \gls{CSMA} fails to provide a reliable channel access. Furthermore, the high user mobility prohibits the use of acknowledgements in \glspl{VANET}, and thereby methods for mitigating the hidden terminal problem. 

The other uncoordinated \gls{MAC} protocols used for \glspl{VC} can be roughly divided into two classes: i) the ones based on \gls{CSMA}, that try to improve its performance by adjusting the load by means of power control~\cite{Moreno09} or transmission rate control~\cite{Huang11}. However, they retain the drawbacks of the original \gls{CSMA}. ii) The second class includes self-organizing protocols predominantly based on \gls{TDMA}, which are  advantageous for overloaded networks~\cite{Bilstrup09}, but cannot guarantee high reliability. These self-organizing algorithms require a learning phase, which increases the channel access delay; moreover, transmission errors during this phase render such protocols unusable.

Recently, a novel uncoordinated \gls{MAC} protocol, called \gls{CSA}, has been proposed for a unicast scenario~\cite{Liva11,Stefanovic13}. It uses the idea of the original slotted ALOHA~\cite{Roberts75} together with \gls{SIC}. The contending users introduce redundancy by encoding their messages into multiple packets, which are transmitted to a \gls{BS} in randomly chosen slots. The \gls{BS} buffers the received signal, decodes the packets from the slots with no collision and attempts to reconstruct the packets in collision exploiting the introduced redundancy. The main difference between \gls{CSMA} and \gls{CSA} it that the former tries to avoid collisions, whereas the latter tries to exploit them.

In this paper, we propose a novel \gls{MAC} protocol for all-to-all broadcast in \glspl{VANET} based on~\gls{CSA}, called \gls{ABCSA}. We consider a scenario where users in the network periodically broadcast messages to the neighboring users. Each user is equipped with a half-duplex transceiver, so that a user cannot receive packets in the slots it uses for transmission\footnote{If full-duplex communication is possible, the analysis of the all-to-all broadcast scenario is identical to that of the unicast scenario.}, which is the main difference of the proposed protocol with respect to the classical unicast~\gls{CSA}. This is modeled as a packet erasure channel and it affects the design and the performance analysis of \gls{CSA}. Using the analytical tools from~\cite{Ivanov14}, we analyze the \gls{PLR} performance of the proposed \gls{ABCSA}. Analytical results show good agreement with simulation results for low to medium channel loads. The results are further used to optimize the parameters of \gls{ABCSA}. Finally, we show that \gls{ABCSA} greatly outperforms \gls{CSMA} for medium to high channel loads.

%Whether \gls{ABCSA} actually works with this \gls{PHY} is left for future investigation.

\section{System Model}\label{sec:syst_model}

%\subsection{Preliminaries}\label{sec:prelim}

There are two types of packets in \glspl{VC}, namely \glspl{DENM} and \glspl{CAM}. \Gls{DENM} packets are sent if requested by a higher-layer application, e.g., in case of an emergency. On the other hand, \gls{CAM} packets are sent periodically every $\tframe$ seconds. In this paper, we consider transmission of \gls{CAM} packets. We assume that all packets have duration $\tpack$, which depends on the packet size $\packsize$ (in bytes), transmission rate $\rdata$, and duration of the preamble added to every packet $\tpream$, i.e., $\tpack = \tpream + \packsize/\rdata$. The parameters in this paper are taken from the \gls{PHY} in~\cite{IEEE80211} and are given in~\tabref{tab:params}.

\begin{table}
\caption{The PHY parameters. The values in the upper part are taken from~\cite{IEEE80211}; the values in the lower part are derived. }
\begin{center}
  \begin{tabular}{l |c| c | c|c}
	  \hline
    Parameter& Variable & \multicolumn{2}{c|}{Value} & Units\\ 
    \hline
    \hline
    Data rate &$\rdata$& \multicolumn{2}{c|}{6}& Mbps\\
	\hline
	PHY preamble &$\tpream$& \multicolumn{2}{c|}{40}& $\mu$s\\
	\hline
	CSMA slot duration &$\tcsma$& \multicolumn{2}{c|}{13}& $\mu$s \\
	\hline
	AIFS time &$\taifs$ & \multicolumn{2}{c|}{58}& $\mu$s \\
	\hline
	\hline
	Frame duration& $\tframe$ & \multicolumn{2}{c|}{100}& ms\\
	\hline
	Guard interval &$\tguard$ & \multicolumn{2}{c|}{5}& $\mu$s\\
	\hline
	Packet size &$\packsize$& 200 & 400& byte\\
	\hline
	Packet duration &$\tpack$& 312 & 576& $\mu$s\\
	\hline
	Slot duration &$\tslot$ & 317 & 581& $\mu$s\\
	\hline 
	Number of slots &$n$ & 315 & 172&\\
	\hline	    
  \end{tabular}
  \end{center}
  \label{tab:params}
\end{table}

\begin{figure}
	\centering
	\includegraphics{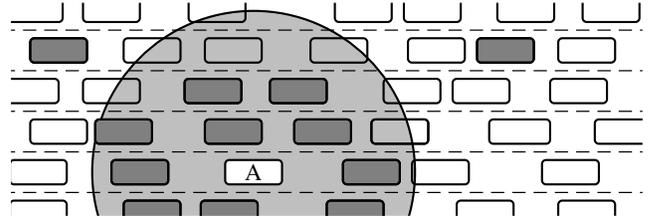}
	\caption{Network model. Rectangles represent users. The circle shows the transmission range of user A. The users marked with gray are neighbors of user~A.}
	\label{fig:network}
\end{figure}

We consider a \gls{VANET} where users are arbitrarily distributed on a 2-dimensional plane as shown in~\figref{fig:network}. Every user has a circular transmission range, e.g., the circle in~\figref{fig:network} shows the transmission range of user A. All users within this transmission range receive packets sent by user A. The transmission range may vary across users, hence, the set of users from which user A receives packets, denoted by $\mathcal{U}$, may be different from the set of users within user A's transmission range . Without loss of generality, we assume that $\mathcal{U}$ consists of $m-1$ users. The users in $\mathcal{U}$ are called the neighbors of user A. As an example, the neighbors of user A are marked with gray in~\figref{fig:network}. We assume that $\mathcal{U}$ remains unchanged during  $\tframe$. In this paper, we focus on the performance of a single user (user A in~\figref{fig:network}), also referred to as the receiver. From the network perspective, the performance of \gls{ABCSA} depends only on the users in $\mathcal{U}$. The rest of the users are ignored as user A cannot receive packets from them.

\subsection{All-to-all Broadcast Coded Slotted ALOHA}

We assume that time is divided into frames of duration $\tframe$. Every user transmits one packet during each frame. Frames are divided into $n = \lfloor \tframe/ \tslot \rfloor$ slots each, where  $\tslot = \tpack + \tguard$ and $\tguard$ is a guard interval that accounts for the propagation delay and timing inaccuracy.  We assume that users are both frame and slot synchronized. This can be achieved by means of, e.g., \gls{GPS}, which provides an absolute time reference for all users.

\begin{figure}
	\centering
	\includegraphics{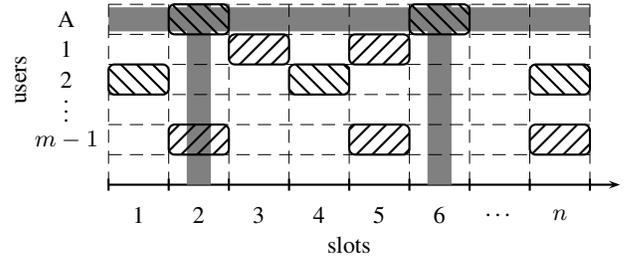}
	\caption{Users' transmissions in B-CSA system within one frame. Rectangles represent transmitted packets. Gray lines indicate the time slots in which user A cannot receive.}
	\label{fig:system_model}
\end{figure}

The transmission phase of the proposed protocol is identical to that of classical \gls{CSA}~\cite{Liva11}, and is briefly described in the following.  Every user maps its message to a PHY packet and then repeats it $l$ times ($l$ is a random number chosen based on a predefined distribution) in randomly and uniformly chosen slots within one frame, as shown in~\figref{fig:system_model}. Such a user is called a degree-$l$ user. Every packet contains pointers to its copies, so that, once a packet is successfully decoded, full information about the location of the copies is available. It is possible to analyze \gls{ABCSA} under idealized assumptions on the \gls{PHY} presented in the following without specifying it explicitly\footnote{We use the \gls{PHY} from~\cite{IEEE80211} mainly to capture timing parameters of the system.}.

Unlike classical \gls{CSA}, where a \gls{BS} is the intended recipient of the messages, every user in \gls{ABCSA} acts as a \gls{BS}, i.e., a user buffers the received signal whenever it is not transmitting. The received signal buffered by user A in slot $i$ is 
\begin{equation}
	y_i = \sum_{j \in \setU_i} h_{i,j} a_{j},
\end{equation}
where $a_j$ is a packet of the $j$th user in $\mathcal{U}$, $h_{i,j}$ is the channel coefficient, and $\setU_i \subset \mathcal{U}$ is the set of user A's neighbors that transmit in the $i$th slot. The $i$th slot is called a \emph{singleton} slot if it contains only one packet. If it contains more packets, we say that a collision occurs in the $i$th slot.

First, user A decodes the packets in singleton slots and obtains the location of their copies. Using data-aided methods, the channel coefficients corresponding to the copies are then estimated. After subtracting the interference caused by the identified copies, decoding proceeds until no further singleton slots are found.

The performance of the system greatly depends on the distribution that users use to choose the degree $l$. The distribution is expressed in the form of a  polynomial,
\begin{equation}\label{eq:distr_orig}
	\lambda(x) = \sum_{l = 0}^{\maxd}\lambda_{l}x^{l},
\end{equation}
where $\lambda_l$ is the probability of choosing degree $l$ and $\maxd$ is the maximum degree.

The performance parameters are defined as follows. The channel load $g = m/n$ shows how ``busy'' the medium around user A is. The average number of users that are not successfully decoded by user A, termed \emph{unresolved users}, is denoted by $\bar{w}$. As reliability is one of the most important requirements in \glspl{VC}, we focus on the average \gls{PLR}, $\PEPt = \bar{w}/(m-1)$, which is the probability of a user to be unresolved, i.e., the probability that its message is not successfully decoded by user A. % fraction of unresolved users, i.e., the users whose messages are not successfully decoded by user A. 

\section{Performance Analysis}\label{sec:analysis}

The typical performance of \gls{CSA} exhibits a threshold behavior, i.e., all users are successfully resolved if the channel load is below a certain threshold value when $n \rightarrow \infty$. The threshold depends only on the degree distribution and is usually obtained via density evolution~\cite{Liva11}. In \glspl{VANET}, however, $n$ is rather small (see~\tabref{tab:params}), which causes an error floor in the \gls{PLR} performance. 

For transmission over a packet erasure channel, it was shown in~\cite{Ivanov14} that the error floor can be accurately predicted based on an \emph{induced} distribution observed by the receiver. As mentioned earlier, in \gls{ABCSA}, a user cannot receive in the slots it uses for transmission. This can be modeled by means of a packet erasure channel. Therefore, the performance of \gls{ABCSA} can be analyzed using the framework in~\cite{Ivanov14}. In this section, we briefly outline the analysis presented in~\cite{Ivanov14} and adapt it to account for the broadcast scenario.

\subsection{Induced Distribution}\label{sec:ind_distr}

Assuming that user A chooses degree $k$, a degree-$l$ user is perceived by user A as a degree-$d$ user if the degree-$l$ user transmits in $l-d$ slots that user A uses for its transmission. From user A's point of view, these $l-d$ slots can be considered erased, which occurs with probability $\binom{n-k}{d}\binom{k}{l-d}/\binom{n}{l}$. Given the constraint $0\le l-d \le k$, the distribution perceived by user A, which we call the $k$-induced distribution, $\lambda^{(k)}(x)$, can be written similarly to~\eqref{eq:distr_orig}, where
\begin{equation}
	\lambda^{(k)}_d = \sum_{l = d}^{\min\{q, k+d\}}   \frac{\binom{n-k}{d}\binom{k}{l-d}}{\binom{n}{l}}\lambda_{l}\label{eq:t_induced}
\end{equation}
is the fraction of users of degree $d$ as observed by user A if it chooses degree $k$. 

It is easy to show that for any finite $k$, $\lambda_l^{(k)} = \lambda_l\,\,\,\forall k,l$ when $n \rightarrow \infty$, i.e., $\lambda^{(k)}(x) = \lambda(x)$ if the number of slots is large enough and $q$ is finite. In this case, density evolution can be used to predict the performance of \gls{ABCSA} in the waterfall region based on the original degree distribution $\lambda(x)$. However, when the number of slots is small, the difference between the original and the induced distributions is significant, especially if $k$ is large. This also suggests that depending on the chosen degree, user A has different decoding capabilities. On the other hand, as shown in~\cite{Ivanov14}, users of different degrees have different protection, namely the higher the degree, the better the protection. We call this property \gls{DUEP}. The rationale behind this property is that the chance of a user to be resolved by other users increases if the user transmits more, but at the same time the chance to resolve other users decreases.

\subsection{Stopping Sets}
In this paper, we focus on the receiving aspect of~\gls{DUEP}, i.e., we characterize different decoding capabilities. The average \gls{PLR} for a degree-$k$ receiver can be defined as
\begin{equation}\label{eq:plr_tot}
\PEPt^{(k)} = \frac{1}{m-1}\sum_{d = 0}^{q}{\bar{w}^{(k)}_d},
\end{equation}
where $\bar{w}^{(k)}_d$ is the average number of unresolved induced degree-$d$ users as observed by a degree-$k$ receiver. For $d = 0$, $\bar{w}^{(k)}_0 = (m-1)\lambda^{(k)}_0$. For other degrees, we show how $\bar{w}^{(k)}_d$ can be approximated in the following. Since user A chooses its degree at random according to $\lambda(x)$, the average \gls{PLR} $\PEPt$ can then be found by averaging~\eqref{eq:plr_tot} over the original degree distribution, i.e.,
\begin{equation}
	\PEPt = \sum_{k = 0}^{\maxd}{\lambda_k \PEPt^{(k)}}. \label{eq:final_plr}
\end{equation}

The only error source in the considered model is the so-called stopping sets. A stopping set is a set of users that cannot be resolved due to an unfortunate choice of slots for transmission. For instance, if two degree-$2$ users transmit in the same two slots they cannot be resolved by any other user. We denote a stopping set by $\setS$ and describe it by a vector $\bm{v}(\setS)=[v_0(\setS), v_1(\setS),\dots, v_q(\setS)]$, where $v_d(\setS)$ is the number of degree-$d$ users in the stopping set $\setS$ and $v_0(\setS) = 0$. For the example above, $\bm{v}(\setS) = [0, 0, 2, 0,\dots, 0]$.

The probabilities of stopping sets can be accurately predicted using the induced distribution in~\eqref{eq:t_induced} and the induced frame length $n^{(k)} = n - k$ as follows. We denote the probability of a stopping set $\setS$ to occur by $\Prss^{(k)}(\setS)$. It can be approximated as~\cite{Ivanov14}
\begin{equation}\label{eq:approximation}
\Prss^{(k)}(\setS) \approx \alpha^{(k)}(\setS) \beta^{(k)}(\setS),
\end{equation}
where $\beta^{(k)}(\setS)$ is the probability of the stopping set $\setS$ to occur calculated under the assumption that $\mathcal{U}$ consists of $v_d(\setS)$ users of degree $d$, $d = 1,\dots, q$. There are, however, more users in $\mathcal{U}$ and this is accounted for by the coefficient $\alpha^{(k)}(\setS)$, defined as the number of combinations to choose $v_d(\setS)$ users out of all degree-$d$ users in $\mathcal{U}$ for $d = 1, \dots, q$. $\alpha^{(k)}(\setS)$ is~\cite{Ivanov14}
\begin{equation}\label{eq:multi_final}
\alpha^{(k)}(\setS) =  \|\bm{v}(\setS)\|_1! \binom{m-1}{\|\bm{v}(\setS)\|_1}\prod_{d = 0}^{q}{\frac{{\left(\lambda_d^{(k)}\right)}^{v_d(\setS)}}{v_d(\setS)!}}.
\end{equation}
$\beta^{(k)}(\setS)$ is not known in general and needs to be found for each stopping set individually based on $\lambda^{(k)}(x)$ and $n^{(k)}$.

Let $\mathcal{A}$ be the set of all possible stopping sets. Using a union bound argument, $\bar{w}^{(k)}_d$ can be upperbounded as
\begin{equation}\label{eq:error_number}
	\bar{w}^{(k)}_d \le \sum_{\setS \in \mathcal{A}}{v_{d}(\setS) \Prss^{(k)}(\setS)} \approx \sum_{\setS \in \mathcal{A}}{v_{d}(\setS) \alpha^{(k)}(\setS) \beta^{(k)}(\setS)}.
\end{equation}
Identifying all stopping sets and calculating the corresponding $\beta^{(k)}(\setS)$ in a systematic way is not straightforward in general. In practice, distributions with large fractions of degree-$2$ and degree-$3$ users are most commonly used since they provide a large threshold~\cite{Liva11}. By running extensive simulations for such distributions, we determined in~\cite{Ivanov14} the set $\mathcal{A}_8$ of eight stopping sets with their $\beta^{(k)}(\setS)$ given in~\cite[eq.~(14) with $n$ replaced by $n^{(k)}$]{Ivanov14} that contribute the most to the error floor. Substituting~\eqref{eq:error_number} with the stopping sets in $\mathcal{A}_8$ into~\eqref{eq:plr_tot} gives an approximation of the average \gls{PLR} for a degree-$k$ receiver.

\subsection{Numerical Results}
We use the analytical error floor approximation~\eqref{eq:final_plr} to optimize the degree distribution for \gls{ABCSA} with finite frame length. We constrain the distribution to only have degrees two, three, and eight to reduce the search space. Such distributions have been shown to have good performance both in terms of error floor and threshold~\cite{Liva11}. The distribution is optimized for the channel load $g = 0.5$ by performing a grid search. The obtained distributions are $\lambda(x) = 0.86x^3 + 0.14x^8$ for $n = 172$ and $\lambda(x) = 0.87x^3 + 0.13x^8$ for $n = 315$.

\begin{figure}
	\includegraphics[]{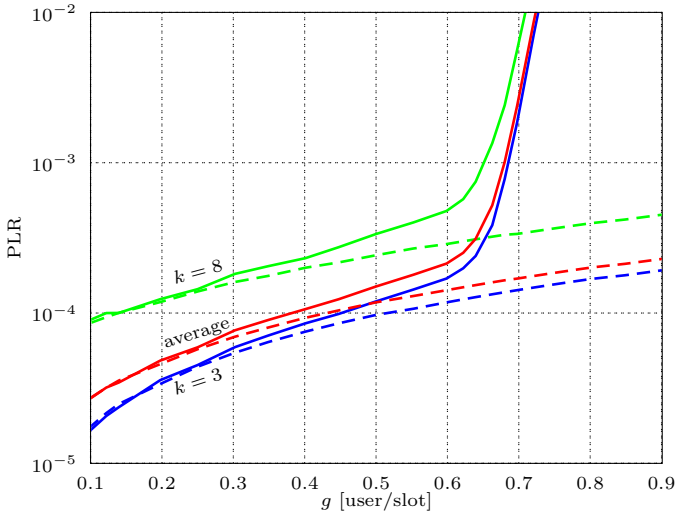}
	\caption{PLR performance of B-CSA for $\lambda(x) = 0.86x^3 + 0.14x^8$ and $n = 172$. The solid lines show simulation results and the dashed lines show analytical approximations~\eqref{eq:plr_tot} and~\eqref{eq:final_plr}.}
	\label{fig:abcsa}
\end{figure}

The analytical error floor predictions in~\eqref{eq:plr_tot} and~\eqref{eq:final_plr} are shown with dashed lines in~\figref{fig:abcsa} for $\lambda(x) = 0.86x^3 + 0.14x^8$ and $n = 172$. Simulation results are shown with solid lines. It can be seen from the figure that the higher the degree of the receiver, the worse the performance. The analytical results show good agreement with the simulation results for low to medium channel loads. In \secref{sec:num_res} we also show that the accuracy improves when $n$ increases.

%To find the average \gls{PLR} of the entire system, degree-$k$ \gls{PLR} $\plr^{(k)}$ needs to be averaged over the original distribution, i.e.,
%\begin{equation}
%	\plr = \sum_{l = 0}^{\maxd}\plr^{(l)}\tilde{\lambda}_{l}. 
%	\label{eq:aver_plr}
%\end{equation}

\section{Carrier Sense Multiple Access}

Comparing \gls{ABCSA} with \gls{CSMA} for the network described in~\secref{sec:intro} is not straightforward for several reasons. First, time is not structured in slots in \gls{CSMA} and new definitions of channel load and \gls{PLR} are therefore needed. This also hinders modeling users' mobility. Second, the behavior of each user in a \gls{CSMA} system depends on the behavior of its neighbors, whereas users in \gls{ABCSA} act independently. Thus, to estimate the performance of an individual user in a \gls{CSMA} system, the entire network needs to be modeled. Third, the performance of \gls{CSMA} for the network in~\figref{fig:network} is affected by the hidden terminal problem since acknowledgements are not used in~\glspl{VANET}. Hence, a proper modeling requires specification of: sensing threshold, path loss, transmitted power, decoding model with signal-to-interference-plus-noise, and actual network geometry. Instead, we introduce a simplified system model which represents the best-case scenario in terms of the performance of \gls{CSMA}. It is easy to implement and compare with \gls{ABCSA}.

We consider a network with $m$ users indexed by $j = 1,\dots, m$, where every user is within all other users' transmission range, i.e., no collision occurs due to the hidden terminal problem. This makes the considered system model favorable for \gls{CSMA} compared to the system model in~\secref{sec:syst_model}. %, as each user can be inhibited by the carrier of each other user, as required by \gls{CSMA}. 
 The set of users is denoted by $\mathcal{V}$. Time is denoted by $t$. At the beginning of contention ($t = 0$), every user chooses a real random number $\tau_j \in [0,\,\,\tframe)$, which represents the time when the $j$th user attempts to transmit its first packet invoking the \gls{CSMA} procedure from~\cite[Fig.~2(a)]{Bilstrup09}. The contention window size $c$ is chosen from the set $\{2^{u}-1 :u \text{ is an interger}\}$, $\taifs$ is the sensing period, and $\tcsma$ is the duration of a backoff slot~\cite{Bilstrup09}. The values of these parameters are specified in~\cite{IEEE80211} (see~\tabref{tab:params} for the values used in  simulations). At time instant $\tau_j + (i-1) \tframe$, the $j$th user attempts to transmit its $i$th packet.  If by this time the $(i-1)$th packet has not been transmitted, the packet is dropped. Each packet is transmitted at most once. 

The channel load is defined as the ratio of the number of users $m$ and $\tframe$ (expressed in slots) to match the definition of the channel load for \gls{ABCSA}, i.e., $g = m/(\tframe/\tslot) = m/n$. The \gls{PLR} for user $j$ is defined as
\begin{equation}
	p_j = \frac{\expect{\tau_1,\dots, \tau_m}{\sum_{\substack{i \in \mathcal{V}\\ i\neq j}} e_i}}{m-1},\label{eq:plr_csma}
\end{equation}
where $e_i \in \{0, 1, 2\}$ is the number of dropped and collided packets of the $i$th user over the time interval $[t_0,\,\, t_0 + \tframe)$ and $\expect{x}{\cdot}$ stands for the expectation over a random variable $x$. For estimating the performance, we introduce a time offset $t_0$ in order to remove the transient in the beginning of contention. As the performance does not depend on the particular user, without loss of generality, we choose user $j$ at random.

\begin{figure}
		\includegraphics[]{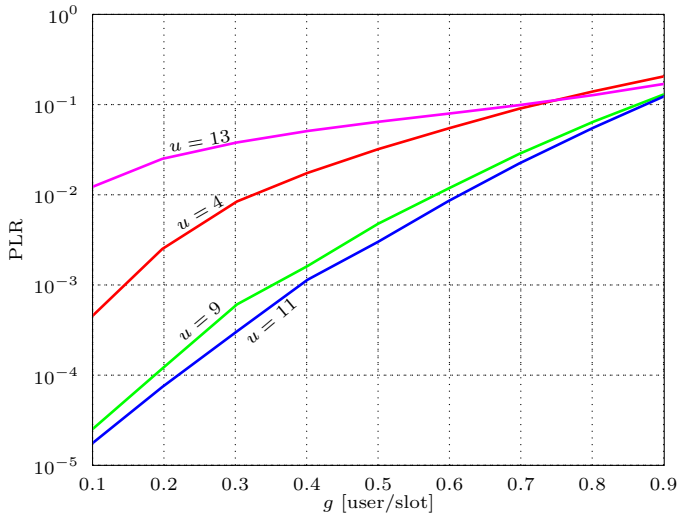}
	\caption{PLR performance of CSMA-CA for 400 byte packets ($n = 172$) and different values of $c = 2^u - 1$ .}
	\label{fig:csma}
\end{figure}

In~\figref{fig:csma}, we show the performance of \gls{CSMA} for different values of the contention window size. We set $t_0 = 2\tframe$ for simulations. Larger values of $c$ reduce the probability of a collision but increase the time required to access the channel: for values of $c \le 2047$ ($u \le 11$), lost packets are only caused by collisions, whereas for $c = 8191$ ($u = 13$), the \gls{PLR} is predominantly defined by dropped packets. It can be seen from~\figref{fig:csma} that the value $c = 2047$ ($u = 11$) provides good performance for the entire range of channel loads. For $u = 10$ and $u = 12$ (not shown in the figure), the PLR performance is slightly worse than that of $u =11$. We remark that the order of the curves remains the same for shorter packets.  $c= 2047$ ($u = 11$) is thus used later on for both packet sizes in~\tabref{tab:params} when comparing \gls{CSMA} with \gls{ABCSA}.

\subsection{B-CSA vs CSMA Comparison}\label{sec:num_res}

\begin{figure}
	\includegraphics[]{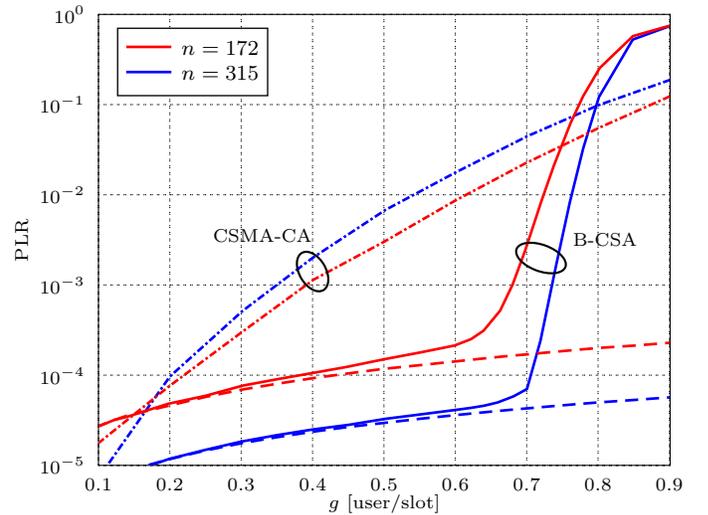}
	\caption{PLR comparison of optimized CSMA-CA and B-CSA for different frame lengths. The solid and the dash-dotted lines show the performance of B-CSA and CSMA-CA, respectively. The dashed lines show the analytical PLR approximation~\eqref{eq:final_plr}. Red and blue lines correspond to $n = 172$ and $n = 315$ slots, respectively.}
	\label{fig:comparison}
\end{figure}

The PLR performance of the two protocols is shown in~\figref{fig:comparison} for $n = 172$ and $n = 315$. The solid and the dash-dotted lines show simulation results for \gls{ABCSA} and \gls{CSMA}, respectively. The dashed lines show the \gls{PLR} approximation~\eqref{eq:final_plr}. The first observation is that the performance of \gls{CSMA} degrades when $n$ increases. This can be explained by the fact that the sensing overhead grows with respect to the packet length when the packet length decreases. On the other hand, the performance of \gls{ABCSA} improves when $n$ grows large, asymptotically approaching the threshold (equal to $0.87$ for the considered distributions) performance predicted by density evolution. This gives an extra degree of freedom when designing a \gls{ABCSA} system, as increasing the bandwidth will decrease the packet duration and, hence, increase the number of slots. We also remark that the accuracy of the analytical \gls{PLR} approximation in~\eqref{eq:final_plr} improves when $n$ increases.

It can be seen from~\figref{fig:comparison} that \gls{ABCSA} significantly outperforms \gls{CSMA} for medium to high channel loads. For example, \gls{ABCSA} achieves a PLR of $10^{-3}$ at channel loads $g = 0.68$ and $g = 0.73$ for $n = 172$ and $n = 315$, respectively. \gls{CSMA} achieves the same reliability at $g = 0.4$ and $g = 0.35$ for $n = 172$ and $n = 315$, respectively, i.e., \gls{ABCSA} can support approximately twice as many users as \gls{CSMA} for this reliability. For heavily loaded networks ($g>0.74$), \gls{CSMA} shows better performance. However, in this case both protocols provide a poor reliability (\gls{PLR} of around $0.1$), which is unacceptable in~\glspl{VANET}.

\section{Conclusions and Future Work}
The proposed protocol can provide an uncoordinated \gls{MAC} all-to-all broadcast with predictable delay and high reliability at large channel loads, which makes it highly appealing for~\glspl{VC}. The proposed \gls{ABCSA} is a cross-layer protocol since the actual \gls{PHY} greatly affects \gls{SIC} performance. Therefore, a more realistic \gls{PHY} needs to be considered. One of the main challenges we foresee here is the channel estimation needed for \gls{SIC}. Among other interesting research directions, we point out the unequal error protection property of \gls{ABCSA} which can be potentially utilized to provide different priorities to packets.

\bibliographystyle{IEEEtran}

\end{document}